\documentstyle[epsf,graphics,a4,12pt]{article}
\begin{document}
\newcommand{\beq}{\begin{equation}}
\newcommand{\eeq}{\end{equation}}
\newcommand{\beqn}{\begin{eqnarray}}
\newcommand{\eeqn}{\end{eqnarray}}
\newcommand{\dpf}{\displaystyle\frac}
\newcommand{\no}{\nonumber}
\newcommand{\ep}{\epsilon}
\begin{center}
{\large Quasinormal modes of Reissner-Nordstr$\ddot{o}$m Anti-de Sitter black holes}
\end{center}
\vspace{1ex}
\centerline{\large Bin
Wang$^{a,b,}$\footnote[1]{e-mail:binwang@fma.if.usp.br},
\ Chi-Yong Lin$^{a,}$\footnote[2]{e-mail:lcyong@fma.if.usp.br}
and Elcio Abdalla$^{a,}$\footnote[3]{e-mail:eabdalla@fma.if.usp.br}
}
\begin{center}
{$^{a}$ Instituto De Fisica, Universidade De Sao Paulo,
C.P.66.318, CEP
05315-970, Sao Paulo, Brazil \\
$^{b}$ Department of Physics, Shanghai Teachers' University,
P. R. China}
\end{center}
\vspace{6ex}
\begin{abstract}
Complex frequencies associated with quasinormal modes for large
Reissner-Nordstr$\ddot{o}$m Anti-de Sitter black holes have been computed. These
frequencies have close relation to the black hole charge and do not
linearly scale with
the black hole temperature as in Schwarzschild Anti-de Sitter case. In terms of AdS/CFT
correspondence, we found that the bigger the black hole charge is, the
quicker for the
approach to thermal equilibrium in the CFT. The properties of quasinormal modes for $l>0$
have also been studied.
\end{abstract} 
\vspace{6ex} \hspace*{0mm} PACS number(s): 04.30.Nk, 04.70.Bw 
\vfill
\newpage

Quasinomal modes of black holes have been an intriguing subject of discussions for the
last few decades. It has become an evidence that the quasinormal ringing will dominate
most processes involving perturbed black holes. This means that quasinormal modes will
carry a unique fingerprint which would lead to the direct identification of the black
hole existence. Detection of these quasinormal modes is expected to be realized through
gravitational wave observation in the near future. In order to extract as much
information as possible from gravitational wave signal, it is important that we
understand exactly how the quasinormal modes behave for the parameters of black holes in
different models. For black holes in asymptotically flat spacetime, especially spherical
cases, they have been studied extensively, for a review see [1]. The study for the
nonspherical black holes is developing [2,3]. Considering the case when the black hole is
immersed in an expanding universe, the quasinormal modes of black holes in de Sitter
space have also been investigated recently [4,5]. It was found that there are qualitative
differences from the asymptotically flat case, in particular the scalar field decay is
always exponential, rather than a power-law tail in asymptotically flat spacetime. This
result has also been uncovered by a very recent study of the Schwarzschild AdS black hole
model [6]. These observations support the earlier argument by Ching et al. [7] that usual
inverse power-law tails as seen in asymptotically flat black hole spacetime, are not a
general feature of wave propagation in curved spacetime.

Motivated by the recent discovery of the AdS/CFT correspondence, the investigation of the
quasinormal modes of AdS black holes becomes more appealing nowdays. The
quasinormal
frequencies of AdS black hole have direct interpretation in terms of the dual conformal
field theory (CFT). In terms of the AdS/CFT correspondence [8-10], a large black hole
corresponds to an approximately thermal state in the field theory, and the decay of the
scalar field corresponds to the decay of perturbation of the state. After computing the
scalar quasinormal modes of Schwarzschild AdS black holes in four, five and seven
dimensions, Horowitz and Hubeny claimed that for large black hole both the real and the
imaginary parts of the quasinormal frequences scale linearly with the
black hole
temperature. The timescale for approaching to the thermal equilibrium is determined by
the imaginary part of the lowest quasinormal frequency and is proportional to the inverse
of the black hole temperature. However, for a small black hole the results are no longer
the same as that for a large black hole. The quasinormal frequencies do not continue to
scale with temperature. Some comments on small black holes were presented therein.

The Schwarzschild AdS black hole studied in [6] is the simplest model in Anti-de Sitter space,
which is determined by only two dimensionful parameters, the black hole
event horizon
$r_+$ and the AdS radius $R$ relating to the cosmological constant by $\Lambda = -
3/R^2$. It is of interest to generalize the study of ref. [6] to a more general model, say
Reissner-Nordstr$\ddot{o}$m (RN) AdS black holes. Besides $r_+$ and $R$, the RN AdS
black hole
has another parameter, the charge $Q$. Thus it possesses richer physics to be explored. In this
paper we are
going to study RN AdS black hole in the hope of getting more understanding of how the
quasinormal modes depend on this additional parameter and whether there is more
information on AdS/CFT correspondence. 

The RN black hole solution of Einstein's equations in free space with a
negative cosmological constant $\Lambda = - 3/R^2$ is given by
\beq    
{\rm d}s^2=-h{\rm d}t^2+h^{-1}{\rm d}r^2+r^2{\rm d}\Omega^2, A=Q/r{\rm
d}t,
\eeq
with
\beq    
h=1-\dpf{r_+}{r}-\dpf{r_+^3}{R^2 r}-\dpf{Q^2}{r_+
r}+\dpf{Q^2}{r^2}+\dpf{r^2}{R^2}.
\eeq
The asymptotic form of this spacetime is AdS. There is an outer horizon
located at $r=r_+$. The mass of the black hole is
\beq          
M=\dpf{1}{2}(r_+ +\dpf{r_+^3}{R^2}+\dpf{Q^2}{r_+}).
\eeq
The Hawking temperature is given by the expression
\beq  
T_H =\dpf{1-\dpf{Q^2}{r_+^2}+\dpf{3r_+^2}{R^2}}{4\pi r_+}
\eeq
and the potential by
\beq 
\phi =\dpf{Q}{r_+}
\eeq
In the extreme case  $r_+, Q$ satisfy the relation
\beq   
1-\dpf{Q^2}{r_+^2}+\dpf{3r_+^2}{R^2}=0.
\eeq

In the following discussions, we will concentrate our attention on the large black hole
with $r_+\gg R$. We will not consider the case of the small black hole here, partly
because it is
unstable, having negative specific heat [11] and partly because it is not of direct
interest for the AdS/CFT correspondence [6].

Let us consider a massless scalar field $\Phi$ in the RN AdS spacetime, obeying the wave
equation
\beq
\Box \Phi =0
\eeq
where $\Box=g^{\alpha\beta}\nabla_{\alpha}\nabla_{\beta}$ is the d'Alembertian operator.
If we decompose the scalar field according to 
\beq    
\Phi=\sum_{lm}\dpf{1}{r}\psi _l (t,r)Y_{lm}(\theta, \phi)
\eeq
then each wave function $\psi _l (r)$ satisfies the equation
\beq       
-\dpf{\partial^2 \psi _l}{\partial t^2}+\dpf{\partial^2\psi _l}{\partial r*^2}=\mho_l\psi
_l
\eeq
where
\beqn       
\mho_l & = & h[\dpf{l(l+1)}{r^2}+\dpf{1}{r}\dpf{dh}{dr}] \no \\
        & = & h[\dpf{l(l+1)}{r^2}+\dpf{r_+ +r_+
^3/R^2+Q^2/r_+}{r^3}-\dpf{2Q^2}{r^4}+\dpf{2}{R^2}]
\eeqn
and $r*$ here is the tortoise coordinate defined by $r*=\int\dpf{dr}{h}$.

The potential $\mho$ has the same characteristic as that in Schwarzschild AdS black
hole. It is positive and vanishes at the horizon, however it diverges at $r=\infty$,
which requires that $\Phi$ vanishes at infinity. This is the boundary
condition to be
satisfied by the wave equation for the scalar field in AdS space. 

Quasinormal modes of AdS space are difined to be modes with only ingoing wave near
the horizon. There only exists a discrete set of complex quasinormal frequencies. 
The quasinormal modes behave like $e^{-i\omega(t+r*)}$ near the horizon in RN
AdS background. We thus introduce the ingoing Eddington coordinates by setting $v=t+r*$. The
metric (1) can be rewritten as
\beq        
{\rm d}s^2=-h{\rm d}v^2+2{\rm d}v{\rm d}r+r^2({\rm d}\theta^2+\sin^2\theta{\rm d}\phi^2)
\eeq
We separare the scalar field in a product form as
\beq     
\Phi=\dpf{1}{r}\psi(r)Y(\theta, \phi)e^{-i\omega v}
\eeq
The minimally-coupled scalar wave equation (7) may thereby be reduced to an ordinary, second
order, linear differential equation with the radial terms yielding
\beq       
h(r)\dpf{d^2\psi(r)}{dr^2}+(h'(r)-2i\omega)\dpf{d\psi(r)}{dr}-V(r)\psi (r)=0,
\eeq
where the potential function is given by 
\beqn   
V(r) & = & \dpf{h'(r)}{r}+\dpf{l(l+1)}{r^2} \no \\
     & = & \dpf{1}{r}(\dpf{r_+}{r^2}+\dpf{r_+^3}{R^2r^2}+\dpf{Q^2}{r_+
r^2}-\dpf{2Q^2}{r^3}+\dpf{2r}{R^2})+\dpf{l(l+1)}{r^2}.
\eeqn
Note that by setting $Q^2=0, R=1$, Eqs.(13,14) go back to the 4D Schwarzschild AdS black
hole case
addressed in [6]. In the following discussion we adopt $R=1$. 

To find the complex values of $\omega$ such that (13) has a solution with $\psi$ finite
at the horizon $r=r_+$, and vanishing at infinity, we have to count on the numerical
calculations. By using the numerical method suggested in [6] to compute the quasinormal
modes, we will expand the solution in power series about the horizon and impose the
boundary condition that the solution vanishes at infinity. Adopting the new variable $x=1/r$,
(13) can be reexpressed as
\beq      
s(x)\dpf{d^2}{dx^2}\psi(x)+\dpf{t(x)}{x-x_+}\dpf{d}{dx}\psi(x)+\dpf{u(x)}{(x-x_+)^2}\psi(x)=0
\eeq
where the coefficient functions are
\beqn       
s(x) & = & \dpf{r_0 x^5-x^4 -x^2 -Q^2 x^6}{x-x_+}    \\
t(x) & = &  3 r_0 x^4 - 2 x^3 - 4 Q^2 x^5 - 2 i \omega x^2  \\
u(x) & = & (x-x_+)V(x)
\eeqn
and the parameter $r_0=\dpf{1+x_+^2 +Q^2x_+^4}{x_+^3}$. As done in [6], we can expand $s,
t$
and $u$ about the horizon $x=x_+$ in the form $s(x)=\sum s_n(x-x_+)^n$ etc. The first
terms are $s_0=2x_+^2\kappa, t_0=2x_+^2(\kappa-i\omega)$ and $u_0=0$, where $\kappa$ is
the surface gravity and has the form $\kappa=(x_+ +3/x_+ -Q^2x_+^3)/2$. The solution of
(15) can then be expressed as a power series
\beq     
\psi(x)=\sum_{n=0}^{\infty} a_n(x-x_+)^n
\eeq

\begin{figure}[htb]
\begin{center}
\leavevmode
\epsfxsize= 8truecm\rotatebox{-90}{\epsfbox{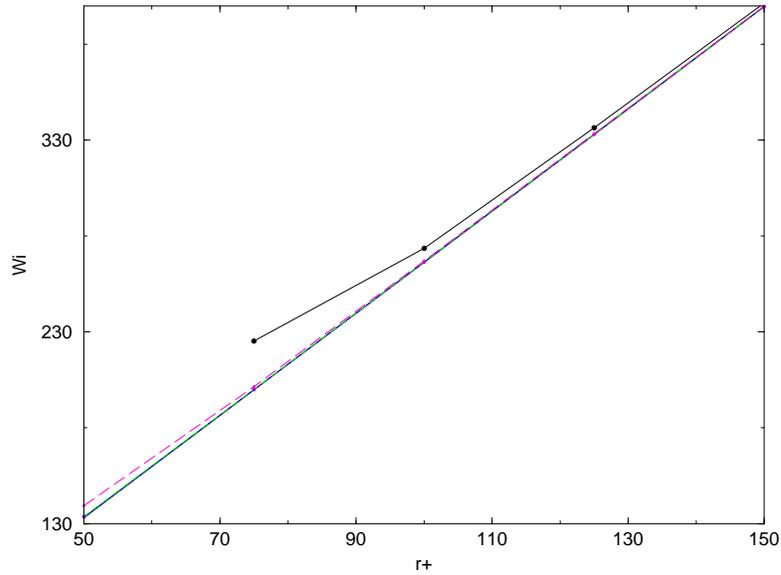}}
\caption{{Lines from the top to the bottom correspond to $Q^2=10^7, 10^6, 10^5$ etc.}}
\end{center}
\end{figure}

\begin{figure}[htb]
\begin{center}
\leavevmode
\epsfxsize= 8truecm\rotatebox{-90}{\epsfbox{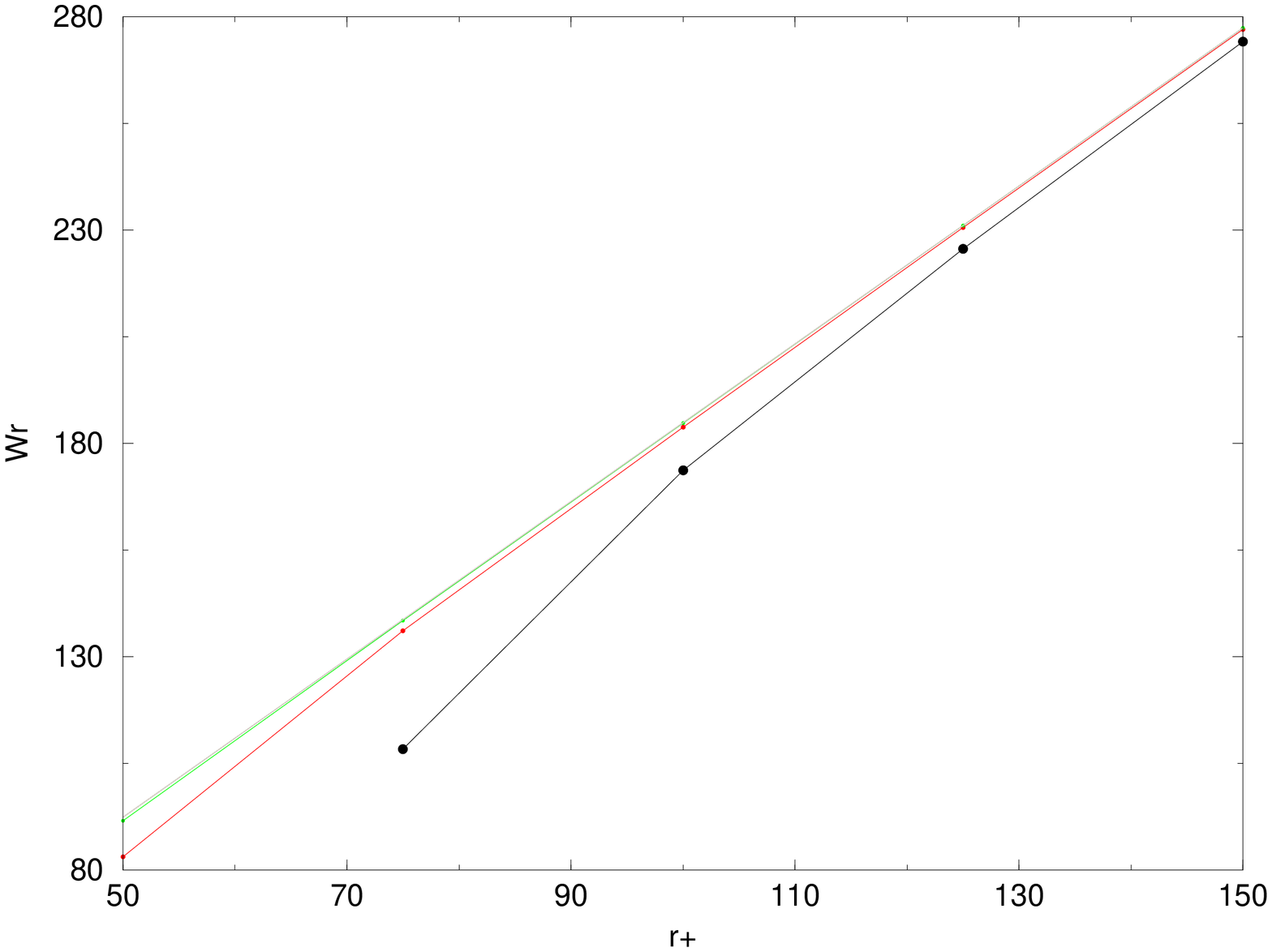}}
\caption{{Lines from the top to the bottom correspond to $Q^2=10^4, 10^5, 10^6, 10^7$
etc.}}
\end{center}
\end{figure}

\begin{figure}[htb]
\begin{center}
\leavevmode
\epsfxsize= 8truecm\rotatebox{-90}{\epsfbox{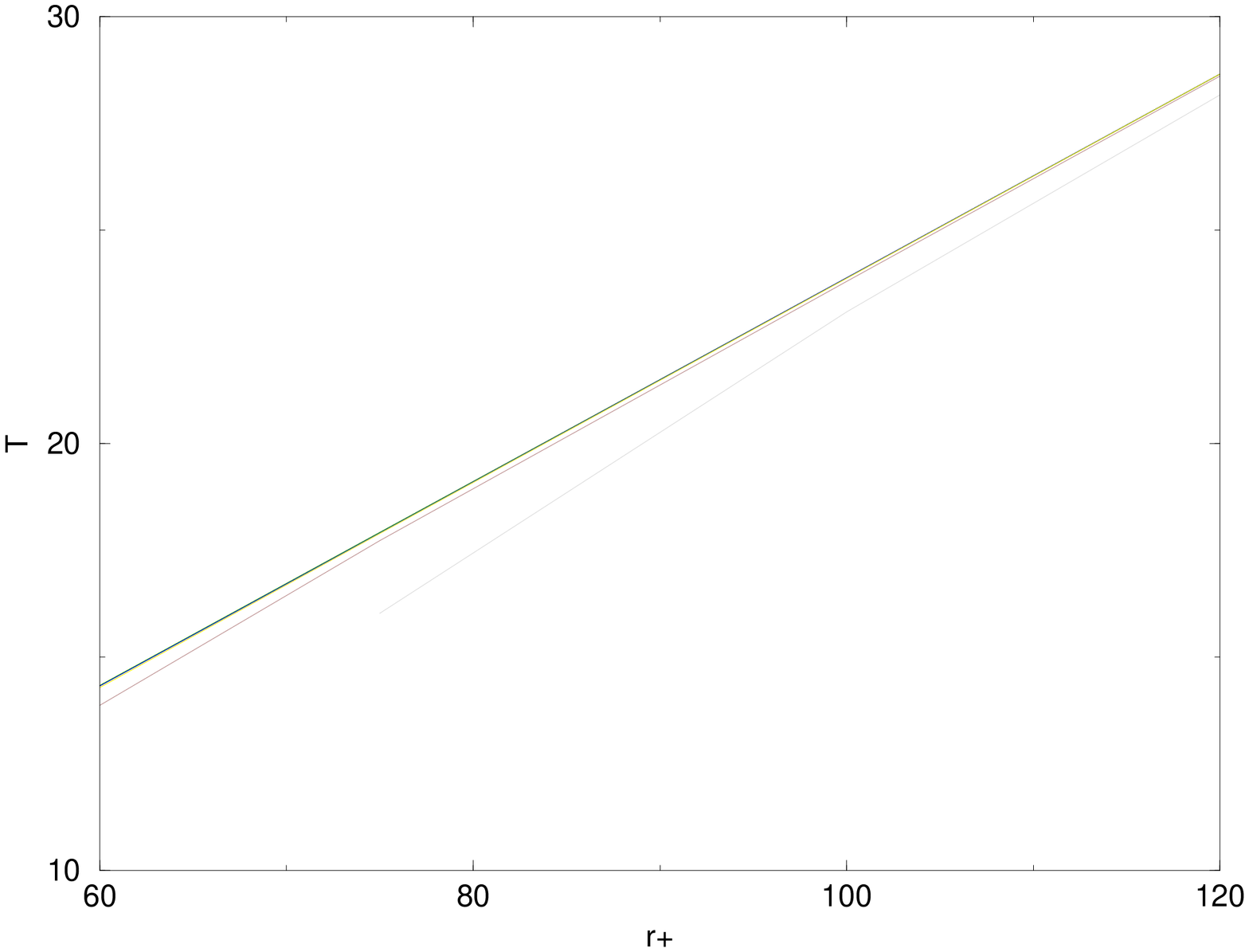}}
\caption{{Lines from the top to the bottom correspond to $Q^2=10^4, 10^5, 10^6, 10^7$
etc.}}
\end{center}
\end{figure}

\begin{figure}[htb]
\begin{center}
\leavevmode
\epsfxsize= 8truecm\rotatebox{-90}{\epsfbox{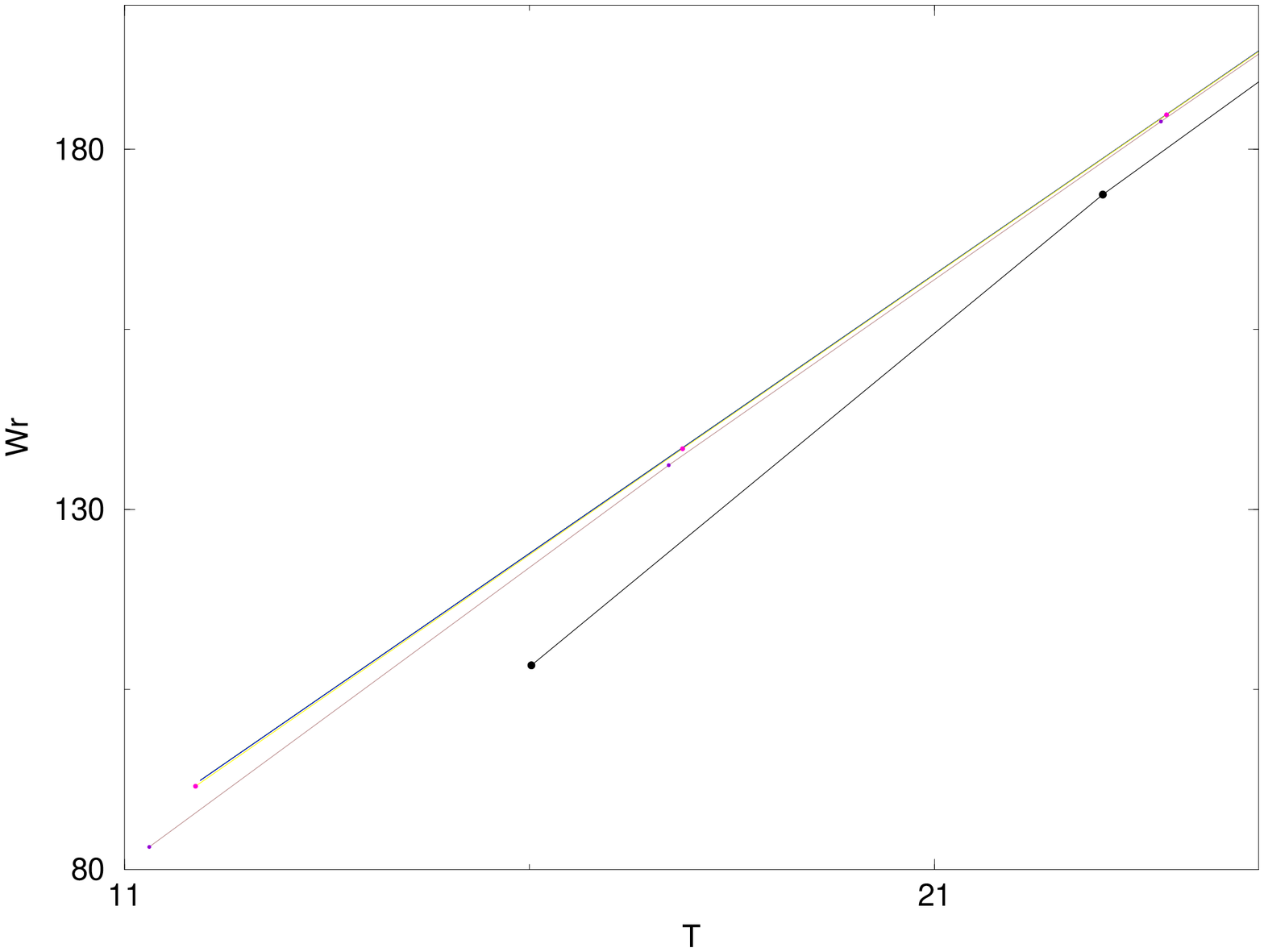}}
\caption{{Lines from the top to the bottom correspond to $Q^2=10^4, 10^5, 10^6, 10^7$
etc.}}
\end{center}
\end{figure}

\begin{figure}[htb]
\begin{center}
\leavevmode 
\epsfxsize= 8truecm\rotatebox{-90}{\epsfbox{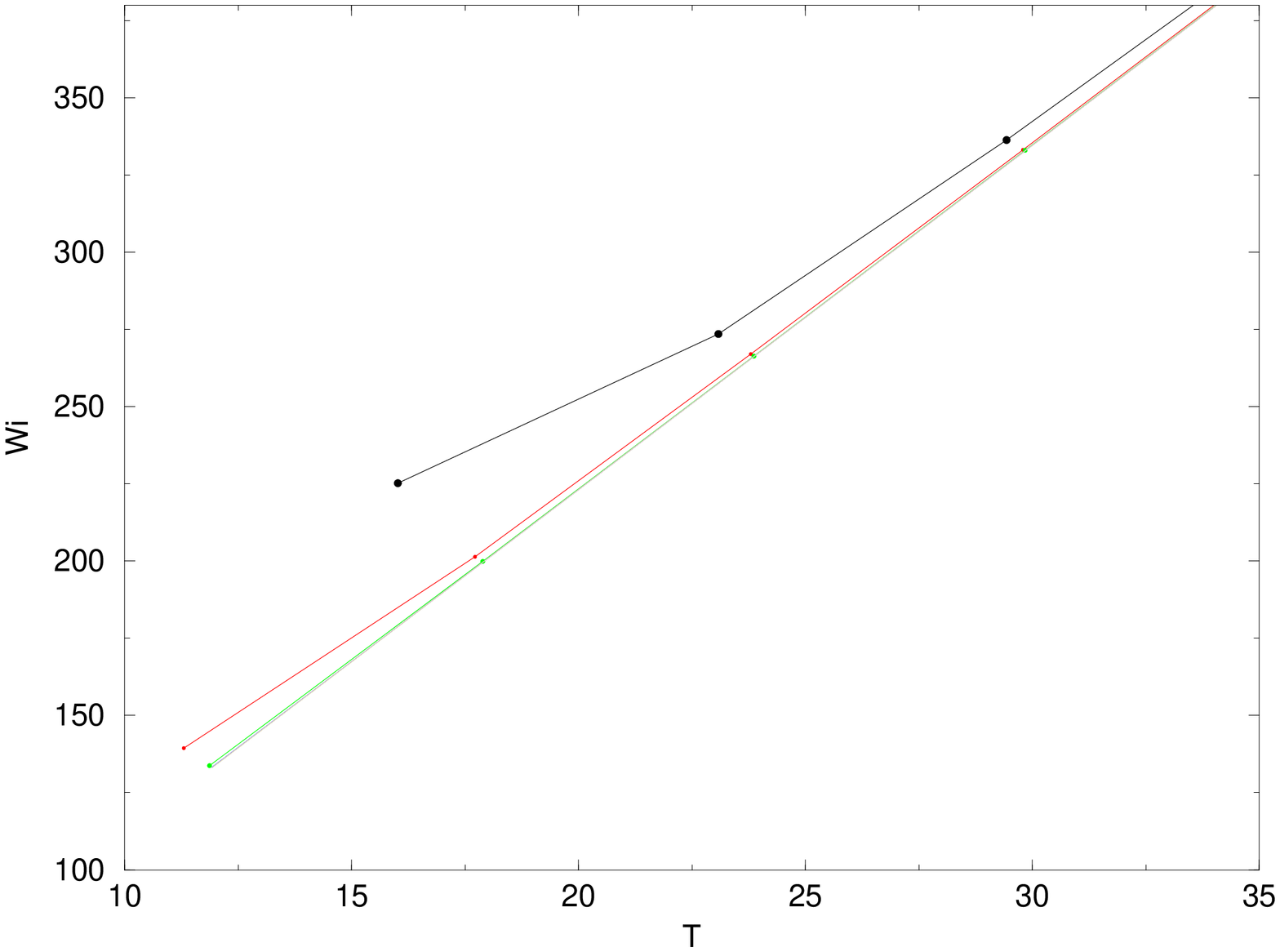}}
\caption{{Lines from the top to the bottom correspond to $Q^2=10^7, 10^6, 10^5$
etc.}}
\end{center}
\end{figure}

\begin{figure}[htb]
\begin{center}
\leavevmode
\begin{eqnarray}
\epsfxsize= 5truecm\rotatebox{-90}{\epsfbox{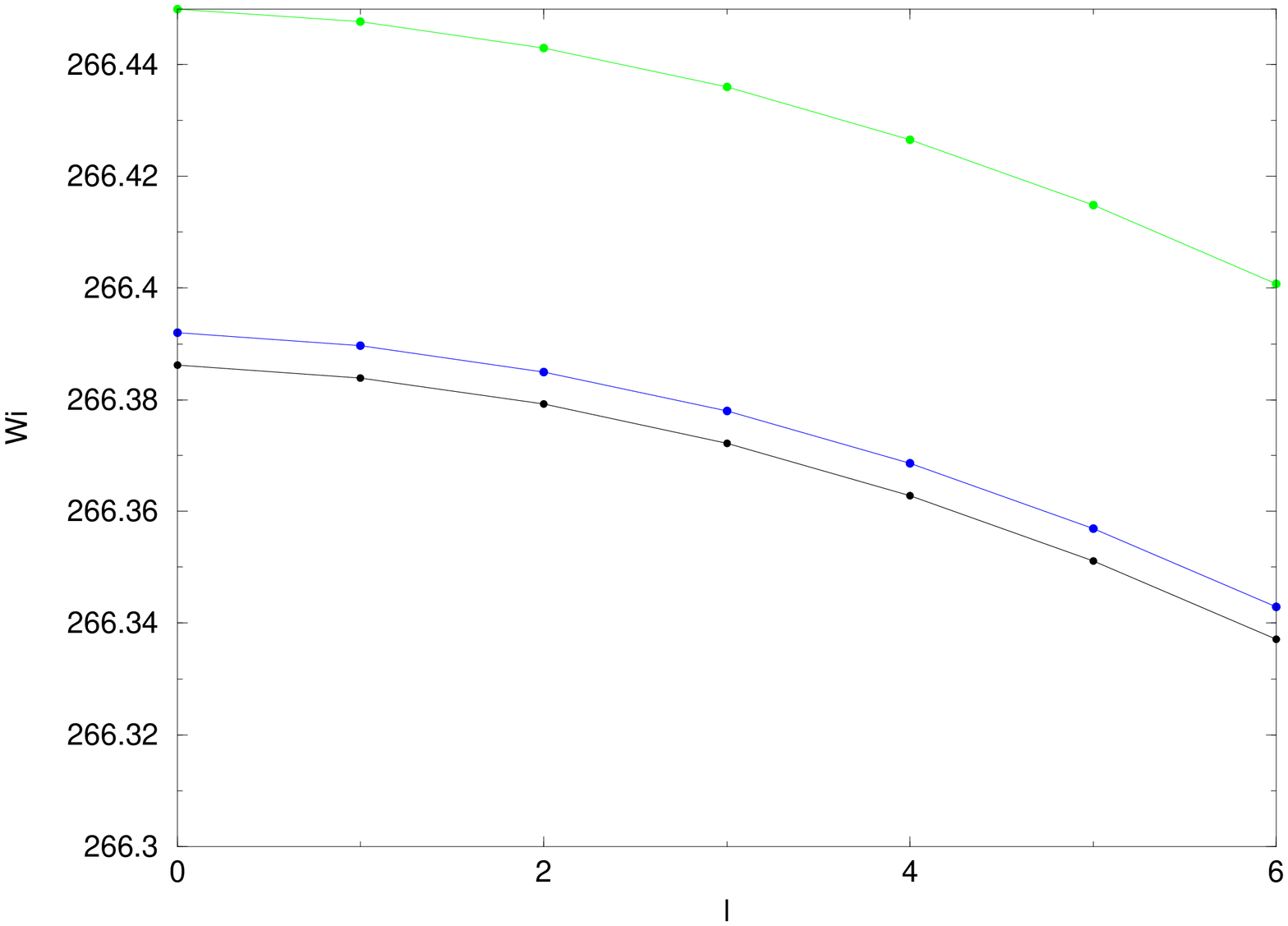}} & &
\epsfxsize=5truecm\rotatebox{-90}{\epsfbox{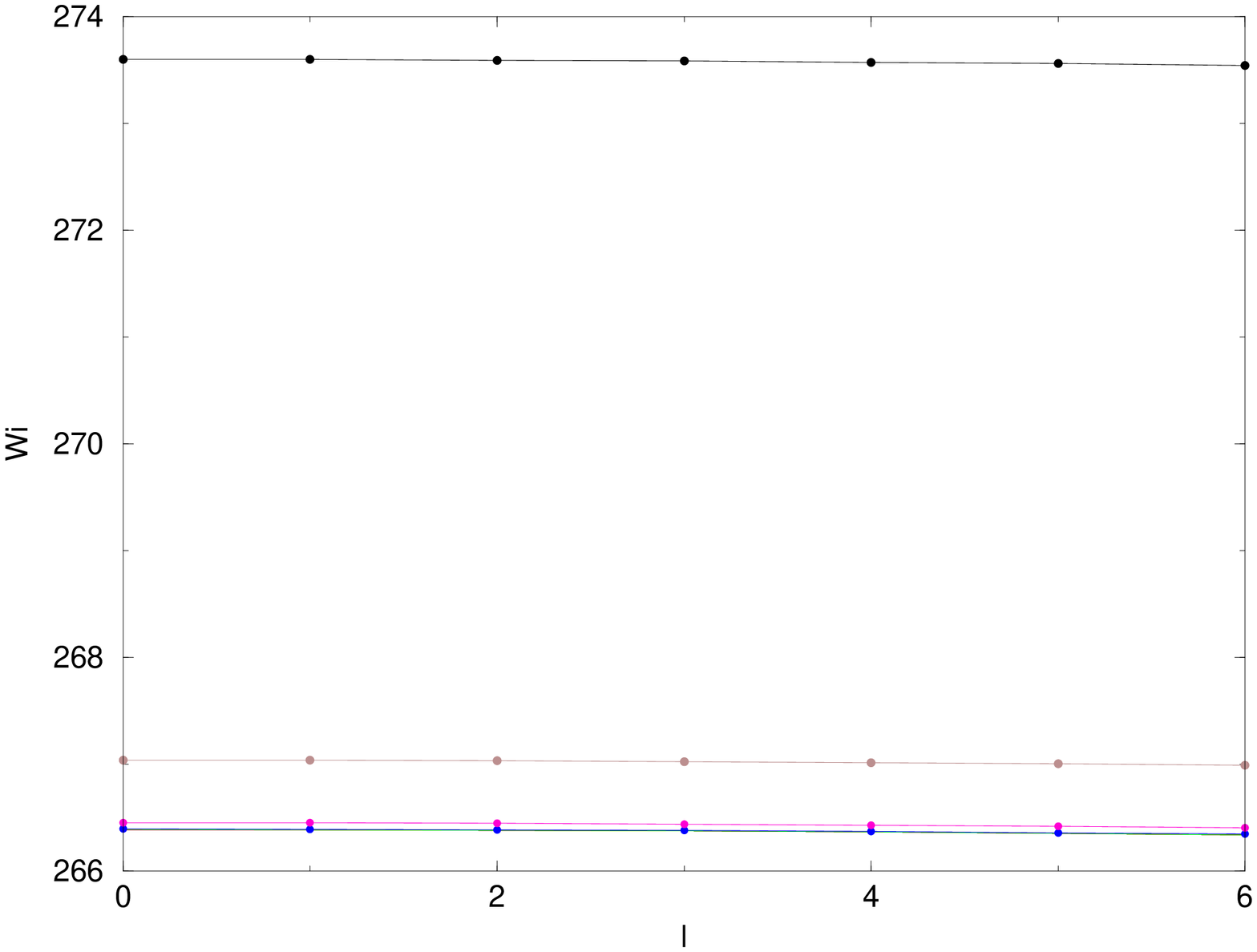}}\nonumber
\end{eqnarray}
\vskip .5cm
\caption{{ Dependence of $\omega_i$ on $l$ for large black hole with
       $r_+=100$. Lines from the top to the bottom on the left figure correspond to 
     $Q^2=10^5, 10^4$ and $10^3$. For the right figure, from the top to the bottom
$Q^2=10^7, 10^6, 10^5, 10^4$.}}
\end{center}
\end{figure}

\begin{figure}[htb]
\begin{center}
\leavevmode 
\epsfxsize= 8truecm\rotatebox{-90}{\epsfbox{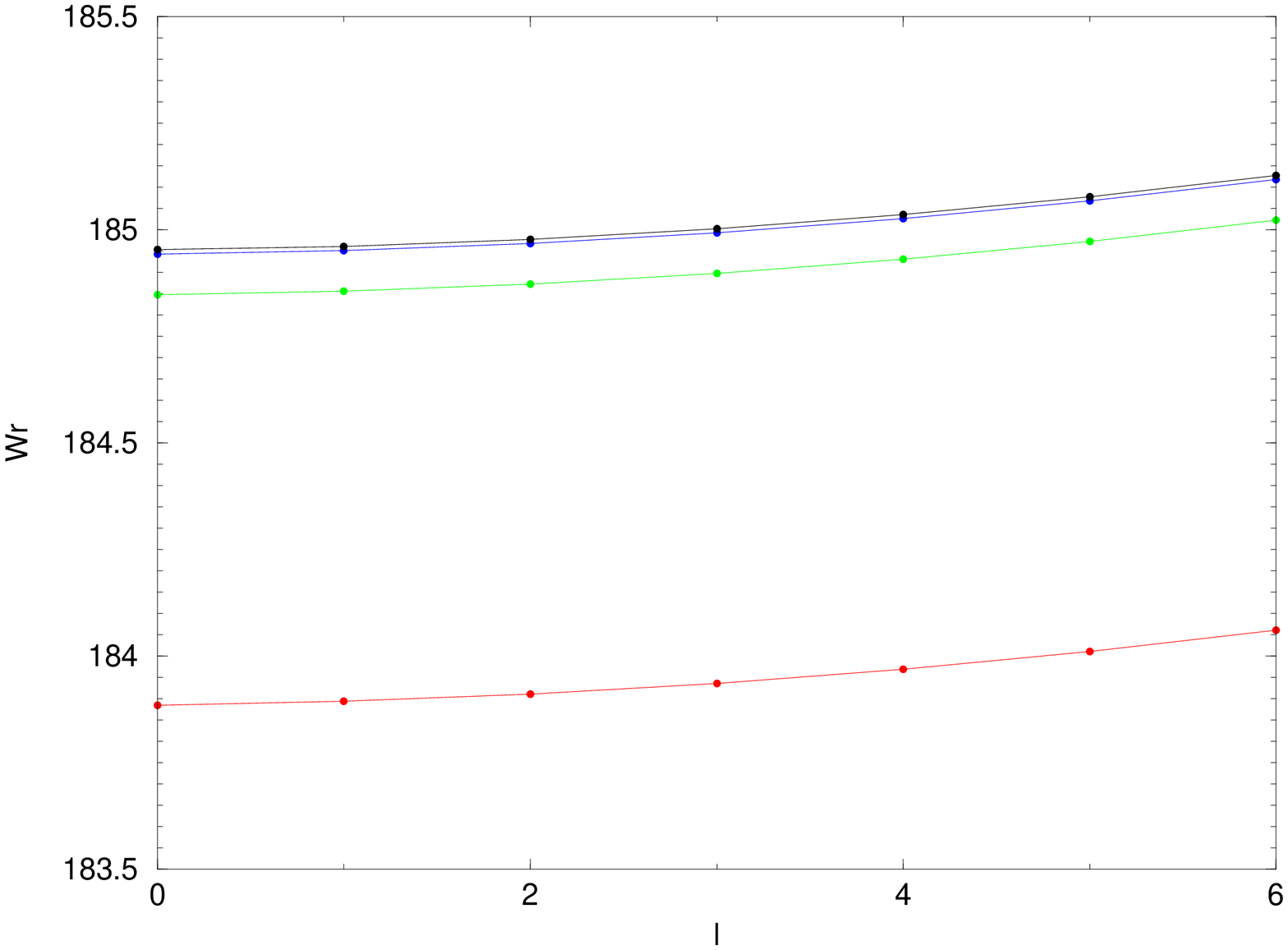}}
\caption{{Dependence of $\omega_r$ on $l$ for large black hole $r_+=100$. Lines from the
top to the bottom correspond to $Q^2=10^3, 10^4, 10^5, 10^6$
etc.}}
\end{center}
\end{figure}

Substituting (19) into (15) and equating coefficients of $(x-x_+)^n$ for
each $n$, we
have the recursion relations for $a_n$
\beq        
a_n=-\dpf{1}{P_n}\sum^{n-1}_{k=0}[k(k-1)s_{n-k}+kt_{n-k}+u_{n-k}]a_k
\eeq
where
\beq     
P_n=n(n-1)s_0+nt_0=2x_+^2 n(n\kappa-i\omega).
\eeq
This relation has the same form as that of the Schwarzschild AdS case, but with an
additional parameter $Q$ herein. 

In order to find the quasinormal modes for the AdS spacetime, we must select solutions
which satisfy the boundary condition $\psi=0$ as $r\rightarrow \infty (x\rightarrow 0)$.
Thus we need to look for the zeros of Eq(19) at $x=0$ in complex $\omega$
planes. For a  given $l, x_+, Q$ the
algorithm to find these frequencies follows the simple steps: 
(i) truncate (19) at a number $N$ of terms, construct the polynormial equation of $\omega$ using
(20,21) so that (19) reduces to
$\sum^N _{n=0} a_n (\omega)(-x_+)^n=0$, (ii) find roots of interest of this function,
(iii)
increase $N$ until these roots become constant  within the desired precision. Since
the problem is to find numerical solutions of a
polynomial equation, it becomes easy to use a built-in Mathematica function to
locate zeros of $\sum ^N _{n=0} a_n (\omega)(-x_+)^n$ directly. This procedure may
reduce tedious
trials and make the numerical calculations neat.

We decompose the quasinormal frequencies into real and imaginary parts in the form
\beq     
\omega=\omega_r - i\omega_i
\eeq
which makes $\omega_i$ positive for all quasinormal frequencies. 
For large black holes $(r_+\gg R)$, the relation of the values of the
lowest quasinomal
mode frequencies for $l=0$ and selected values of $r_+$ for different charge $Q$ are
exhibited in fig.1 and fig.2. The dots represent the lowest modes. In fig.1, lines from
the top to the bottom correspond to $Q^2=10^7, 10^6, 10^5$ etc. However the lines shown
in fig.2 are corresponding to $Q^2=10, 10^2,..10^7$ from the top to
the bottom, respectively. It is easy to see that with an additional parameter, the charge
$Q$, neither the real nor the imaginary part of the frequency is a linear function of
$r_+$ as found in Schwarzschild AdS case [6]. The bigger the charge $Q$ is, the larger
is the deviation from the linear relation we observe. 

The imaginary  and real parts of the quasinormal frequencies relate to the damping
time scale $(\tau_1=1/\omega_i)$ and oscillation time scale $(\tau_2=1/\omega_r)$,
respectively. From fig.1 we learn that as $Q$ increase, $\omega_i$ increases as well,
which corresponds to the decrease of the damping time scale. According to the AdS/CFT
correspondence, this means that for big $Q$, it is quicker for the quasinormal ringing to
settle down to the thermal equilibrium. Fig.2 tells us that the bigger charge $Q$ leads
to the smaller $\omega_r$, which means that the frequency of the oscillation becomes
small as $Q$ increases. Therefore from fig.1 and fig.2 we can have a picture that if we
perturbe a RN AdS black hole with high charge, the surrounding geometry  
will not ``ring"
as much and long as that of the black hole with small $Q$. It is easy for the
perturbation on the highly charged AdS black hole background to return  to
thermal
equilibrium. This is the new physics brought by the additional parameter $Q$ in RN AdS
black hole.

>From (4), we learn that the temperature of the large charged AdS black hole does not
scale linearly with the event horizon as that in Schwarzschild AdS black hole case. The
behavior of the temperature is shown in fig.3. The linear relation between $T$ and $r_+$
is broken as $Q$ increases. The relations between the real and the imaginary parts of the
frequency and the temperature for large RN AdS black hole are displayed in fig.4,5. Again
we see that in contrast to the results in [6], the relations are no longer linear when
the additional parameter $Q$ is taken into account.

We have so far discussed only the lowest quasinormal modes with $l=0$. Increasing $l$, we
obtain the surprising effect of increasing the damping time scale ($\omega_i$ decreases),
and decreasing the oscillation time scale ($\omega_r$ increases) as in the case of
Schwarzschild AdS black holes. As pointed out in [6], here we may also meet the problem
of possible negative $\omega_i$ as it continuously decreases with $l$.
However from fig.6,
it looks that for the large black hole, the problem is not as serious as that shown in
[6] for small Schwarzschild AdS case. Despite the similar behavior, once
again the
additional parameter $Q$ led us to further new properties. As fig.6 shows, different
values of $Q$ do not change the qualitative characteristic of decreasing $\omega_i$ with
$l$. Their decreasing rates are the same and do not depend on the value $Q$. However it
is clear that the higher is the charge of AdS black hole, the later we will confront
the tough question of $\omega_i$ as $l\rightarrow \infty$. The dependence of $\omega_r$
on $l$ for different values of $Q$ are exhibited in fig.7. Lines from upper to the bottom
are $Q^2=10^3, 10^4...10^6$, respectively. As shown for the case of the
imaginary part, the lines of
different charges are parallel.

In our numerical calculations, we found that we need for a large number $N$
of terms in the partial sum
to reduce the relative error in the computation of quasinormal frequencies as $r_+$
decreases and $Q$ increases. However when the charge $Q$ increases to nearly the extreme
value satisfying (6), we cannot accurately determine the mode, no matter how
large is the value of $N$ that we
adopt. The numerical convergence problem can be attributed to the method
we adopted to
compute the quasinormal modes. In the expansion of the solution of the differential
equation into power series, we have the convergence radius
\beq       
L=\lim_{n\rightarrow
\infty}\vert\dpf{\psi_n}{\psi_{n+1}}\vert \sim \vert\dpf{3+x_+^2-Q^2x_+^4}{9-3x_+ +
4x_+^2
-x_+^3 -Q^2(5x_+^4 -x_+^5)}\vert
\eeq
To expand the solution in a power series about the horizon, we need
$L>x_+$ to ensure
that the expansion is valid. In the range $0<Q^2<\dpf{9-3
x_+ +4x_+^2-x_+^3}{5x_+^4-x_+^5}$, $L$ increases from $L_0=(3+x_+^2)/(9-3x_+
+4x_+^2-x_+^3)$
to infinity, where $L_0>x_+$ for a big black hole, which means that $L>x_+$ always hold in
this range. However for $Q^2>\dpf{9-3
x_+ +4x_+^2-x_+^3}{5x_+^4-x_+^5}$, $L$ continuously decreases with increasing values of $Q$.
Finally when $Q^2\rightarrow
\dpf{3+x_+^2}{x_+^4}=3r_+^4+r_+^2$, which is the exact extreme value, $L\rightarrow
0<x_+$.
Therefore although the
numerical approach is very efficient for determining the quasinormal modes for
Schwarzschild AdS black hole and lowly charged AdS black hole, it breaks down for
the nearly extreme RN AdS black hole case. 

It is important to notice that Eq(15) has the same form as the generalized spheroidal
wave equations which also arises
both in the quantum scattering theory of nonrelativistic electrons from polar molecules
and ions and in the theory of radiation process involving black hole in asymptotically 
flat spacetime. Similar convergence problem also appeared and hindered the
earlier study of
the quasinormal modes for extreme charged black hole and extreme rotating black hole
[11,12]. 

Besides problems related to the method, is there some  deep physics to account
for the similar no
convergence problem in determining quasinormal modes for extreme black
holes both in
asymptotically flat and AdS spaces? We know that the frequencies and
damping times of the quasinomal modes are entirely fixed by the black hole, and are
independent of the initial perturbation. It has been shown that there is a second order
phase transition in the extreme limit of black holes [14-16]. This result
has been
further promoted in a recent study for the charged AdS black holes [11,17]. Because of
the phase transition, the fluctuation of thermodynamic quantities become tremendous. It
is hard to expect that we can obtain the fixed quansinormal frequencies to
characteristize the thermalization timescale in the strong coupled CFT on the extreme RN
AdS background. We speculate that the black hole phase transition maybe the candidate
physical reason behind this problem.

In summary, we have computed the scalar quasinormal modes of large RN AdS black hole.
These modes govern the late time decay of a minimally coupled scalar field, such as the
dilaton. Compared to the Schwarzschild AdS black hole, we found that the additional
parameter $Q$ in RN AdS black hole has brought in some new properties. We observed that
these modes no longer linearly scale with the black hole temperature. By AdS/CFT
correspondence, we can interpretate the decay of the quasinormal modes to the time scale
approaching thermal equilibrium in CFT. We learnt that perturbations are associated with the
black hole charge. The larger the charge of the RN AdS black hole, the sooner it returns to
the thermal equilibrium. The dependence of the quasinormal frequencies on nonzero angular
momentum $l$ has also been discussed. Due to the convergence problem, the method
we adopted cannot be extended directly to study of the extreme black holes cases. Further
refinement of the numerical method of solving the differential equation (13) is called
for. Based upon some critical phenomena uncovered, we have given some speculation of the
possible physical reason behind the problem for the extreme black holes.

ACKNOWLEDGMENT: 
This work was partially supported
by Fundac$\tilde{a}$o de Amparo $\grave{a}$ Pesquisa do Estado de
S$\tilde{a}$o Paulo (FAPESP) and Conselho Nacional de Desenvolvimento Cientifico e Tecnologico (CNPq).
B. Wang would  like to acknowledge the support given by Shanghai Science
and Technology Commission.

\end{document}